\documentclass[aps,preprint,amsmath,amssymb,showpacs,amsfonts,nofootinbib]{revtex4}
\usepackage{epsfig}
\usepackage{graphicx}
\usepackage{dcolumn}
\usepackage{bm}
\usepackage{color}

\newcommand{\beq}{\begin{equation}}
\newcommand{\eeq}{\end{equation}}
\newcommand{\bea}{\begin{eqnarray}}
\newcommand{\eea}{\end{eqnarray}}

\def\d{\partial}

\def\d{\delta}

\def\phi{\varphi}

\def\l{\lambda}
\def\m{\mu}
\def\n{\nu}

\def\p{\pi}

\def\s{\sigma}

\def\th{\theta}

\def\nab{\nabla}

\begin{document}

\title{The universal viscosity to entropy density ratio from entanglement}

\author{Goffredo Chirco}
\author{Christopher Eling}
\author{Stefano Liberati}

\affiliation{SISSA, Via Bonomea 265, 34136 Trieste, Italy and INFN Sezione di Trieste, Via Valerio 2, 34127 Trieste, Italy}

\begin{abstract}

We present evidence that the universal Kovtun-Son-Starinets shear viscosity to entropy density ratio of $1/4\pi$ can be associated with a Rindler causal horizon in flat spacetime. Since there is no known holographic (gauge/gravity) duality for this spacetime, a natural microscopic explanation for this viscosity is in the peculiar properties of quantum entanglement. In particular, it is well-known that the Minkowski vacuum state is a thermal state and carries an area entanglement entropy density in the Rindler spacetime. Based on the fluctuation-dissipation theorem, we expect a similar notion of viscosity arising from vacuum fluctuations. Therefore, we propose a holographic Kubo formula in terms of a two-point function of the stress tensor of matter fields in the bulk. We calculate this viscosity assuming a minimally coupled scalar field theory and find that the ratio with respect to the entanglement entropy density is exactly $1/4\pi$ in four dimensions. The issues that arise in extending this result to non-minimally coupled scalar fields, higher spins, and higher dimensions provide interesting hints about the relationship between entanglement entropy and black hole entropy.

\end{abstract}

\maketitle

\section{Introduction}

One of the few low energy artifacts of quantum gravity we currently possess is that the combination of the gravitational field equations and quantum field theory require black holes to behave as thermodynamic objects. They have an entropy proportional to the cross-sectional area of the event horizon divided by the square of the Planck length \cite{Bekenstein:1973ur} and a temperature due to quantum
Hawking radiation \cite{Hawking:1974sw}. The subsequent development of the black hole {\it membrane paradigm} \cite{Damour, Price, membrane} showed that this behavior also extends to non-equilibrium thermodynamics. The equations governing dynamical horizons seem to be those of a viscous fluid, with apparent hydrodynamic transport coefficients such as viscosities. In general though, the relationship between the dynamics of a fluid and the dynamics of any black hole event horizon is just an analogy.  The reason is that hydrodynamics is only a valid effective theory of many-body systems on long spatial and time scales \cite{Forster,Landau10}.
%In order for hydrodynamics to be a valid description, the characteristic wavelength and time scale of perturbations to the system must be much larger than the microscopic scale set by a correlation length (or mean free path).
This basic criterion cannot be fulfilled even in the familiar example of a spherically symmetric Schwarzschild horizon.
This is the reason why the membrane paradigm relates the black hole horizon to a fictitious fluid with unphysical negative bulk viscosity \cite{Eling:2009pb,Eling:2009sj}.

However, there are black hole spacetimes where a large scale hydrodynamic limit exists. Important examples are black holes and branes in asymptotically Anti-de Sitter (AdS) spacetimes. These have been extensively studied in the literature over the past decade due to their role in the celebrated AdS/conformal field theory (CFT) correspondence \cite{Maldacena:1997re,Aharony:1999ti}. The correspondence relates (quantum) gravity in $(D+1)$-dimensional asymptotically AdS spacetimes to certain conformal field theories on the $(d+1)$-dimensional AdS boundary with $d=D-1$. In the duality, a classical black hole in AdS spacetime corresponds to a strongly coupled thermal CFT on the boundary at the Hawking temperature. The large scale dynamics of the black hole therefore is dual to the hydrodynamics of the thermal gauge theory \cite{fluidgrav}.

Hydrodynamic transport coefficients such as viscosities are calculated from a microscopic theory using ``Kubo formulas", which involve finite temperature Green's functions of conserved currents. This is not an easy calculation even at weak coupling (see for example, \cite{Jeon:1994if}), and seems to be extremely hard at strong coupling. However, the duality picture allows one to determine the transport coefficients of these strongly coupled theories in a fairly straightforward way by mapping the calculation of Green's functions into a classical boundary value problem in the bulk spacetime \cite{Son:2007vk}. An application of this mapping is that the transport coefficients of the dual gauge theory can be calculated directly at the black hole horizon from the membrane paradigm \cite{dualitymembrane,Iqbal:2008by}. A key early result that emerged from this work is that, in the limit of infinite coupling, any (not necessarily conformal) gauge theory with an Einstein gravity dual has a shear viscosity to entropy density ratio of $\eta/s = \hbar/4\pi k_B$. This value was conjectured by Kovtun, Son, and Starinets (KSS) to be a universal lower bound \cite{Kovtun:2004de}. Using the membrane formalism, general formulas have been recently developed which characterize the shear viscosity of gauge theories with generalized gravity duals in terms of an effective coupling of gravitons at the horizon \cite{effcouplings}.

Although the universal KSS ratio seems to be rooted in gravitational physics, curiously it does not depend on the Newton constant $G_N$.
Furthermore, the ratio also appears to be {saturated} even for a Rindler acceleration horizon in flat Minkowski spacetime \cite{RindlerKSS}, where gravity is absent. Indeed, one can assume that, like a black hole, the Rindler causal horizon can be endowed with a finite area entropy density $s$. Although there is no holographic duality like AdS/CFT in this case, the hydrodynamic limit exists and a shear viscosity of $\hbar s/4\pi k_B$ emerges when one studies the dynamics of the horizon using the membrane paradigm \cite{Eling:2009pb, Eling:2009sj}.

However, in the absence of a clear holographic duality, the interpretation of this shear viscosity to entropy density ratio seems to be unclear. For example, what is the underlying fluid system that is being probed by these calculations?

The fluctuation-dissipation theorem links viscous dissipation to fluctuations of a thermal equilibrium state. An attempt to interpret the viscous dissipation rate of a horizon in terms of the quantized gravitational fluctuations of the horizon shear {was already developed many years ago \cite{Candelas:1977zz}}.  Here we take a different approach, based on the notion of quantum entanglement together with the properties of vacuum fluctuations.

It is well-known that observables restricted to the Rindler ``wedge" of the global spacetime perceive the Minkowski vacuum to be a mixed thermal state at the Tolman-Unruh temperature \cite{Unruh:1976db}. In addition, there is a corresponding statistical entanglement entropy for matter fields in the Rindler wedge. This quantity is quadratically ultraviolet (UV) divergent, due to the infinite redshift/blueshift at the horizon. When a cut-off is introduced, the entropy scales not with volume of the wedge, but instead like the area of the horizon boundary. Hence the Rindler wedge is equipped with thermodynamic properties, which seem to be naturally encoded into a ``pre-holographic" lower dimensional description associated with the horizon boundary.

On large scales this thermal vacuum state should behave as a fluid, with hydrodynamics as an effective description. In this regime, we expect to find a holographic ``entanglement viscosity" which, when similarly cut off, scales exactly with the entanglement entropy so that the KSS ratio is satisfied universally. To test this hypothesis, we propose a microscopic Kubo-like formula for the shear viscosity associated with the fluid description of the vacuum thermal state. The Kubo formula is constructed from the Green's functions of the energy-momentum stress tensor for the matter fields in the wedge. All quantum fields in nature must contribute to the vacuum fluctuations and therefore to the entanglement entropy and viscosity. For simplicity, we start by considering a free, minimally coupled scalar field theory.  Remarkably, we show that the ratio of our shear viscosity to the entanglement entropy density is exactly the KSS ratio. This suggests that the KSS ratio may be a fundamental holographic property of spacetime (rather than just of the aforementioned AdS black hole solutions).

However, demonstrating the equivalence of the above defined ratio with the universal KSS one would require extending it to different field theories and higher dimensions.We try to generalize our approach with the simplest extension to a non-minimally coupled scalar field theory. However, while the viscosity appears to be independent of the coupling to the scalar curvature, the status of the entropy density in the literature is less clear, the issue being closely tied to the general relationship between entanglement entropy and the black hole entropy. Finally, we re-calculate the ratio in a higher dimensional spacetime. Here the $1/4\pi$ result is apparently not readily extendable and  we shall discuss the possible reasons why more work is needed in this direction (and the possible insights this investigation might lead to).

This paper is structured as follows. In Section II, we review the thermal properties of the Rindler wedge and the notion of entanglement entropy. In Section III, we discuss two examples where a shear viscosity emerges from classical hydrodynamics applied to the Rindler thermal state. This serves as a motivation for the Kubo formula developed in Section IV.  Section V contains our calculations for the free, non-minimally coupled scalar field. We conclude in Section VI with the possible implications of our result,  a discussion of the relationship between entanglement entropy and black hole entropy, and extensions to higher dimensions.

\section{The Rindler wedge and entanglement}
\label{Rindler}

We start by reviewing the properties of the Rindler wedge associated with an accelerated observer in Minkowski spacetime. Consider a general $(D+1)$-dimensional flat manifold in the Rindler coordinates $Y^A = (\tau, \xi, x^i)$, where $i=1..d$ ($d$ is the number of transverse spatial dimensions),
\beq ds^2 = g_{AB} dY^A dY^B = \kappa^2 \xi^2 d\tau^2- d\xi^2 - \sum^{d}_{i=1} dx^i dx_i. \label{Rindlermetric} \eeq
Here $\kappa$ is an arbitrary constant with dimensions $[L]^{-1}$ (we now work in units where $\hbar=c=k_B=1$) associated with the normalization of the timelike Killing vector $\partial_\tau$.  The Rindler metric can be obtained by a coordinate transformation
of the usual Minkowski inertial coordinates $X^A$,
\bea t &=& \xi \sinh (\kappa \tau)\nonumber\\
z &=& \xi \cosh(\kappa \tau)\nonumber\\
x^i &=& x^i \label{transform}.
\eea
%
%Lines of constant $\xi$ are the hyperbolas
%$t^2-z^2=\xi^2$, which are the worldlines of accelerating observers with proper acceleration $a = \xi^{-1}$.
Unlike the global inertial coordinates $X^A$, the Rindler coordinates only cover a ``wedge" subregion of Minkowski space where $z > |t|$. The timelike Killing flow $\partial_{\tau}$ is equivalent to a continuous boost in the $z$ direction. The respective boost time parameter  $\tau$ is proportional to the proper time along the wordlines of the uniformly accelerated observer, defined by the $\xi=const$ hyperbolas. The null surface $z=t$ acts just like the future event horizon of a black hole since the points ``inside" are causally disconnected from the accelerated observers.

The Rindler spacetime therefore mimics many of the properties of black holes in curved spacetimes. This statement is true not only at the classical level, but also when quantum effects are included. In the 1970's it was realized that quantization of fields on the Rindler spacetime (\ref{Rindlermetric}) is inequivalent to the usual field quantization in full Minkowski spacetime. The reason is that the Rindler Hamiltonian generates a flow in boost time. It follows that the notion of the vacuum for the Rindler quantization must be different than the usual Minkowski vacuum $|0  \rangle $.  A non-inertial observer will determine a different zero energy state, called the Fulling-Rindler vacuum $|F  \rangle $ \cite{Fulling:1972md}. Unruh's discovery \cite{Unruh:1976db} was that the ordinary Minkowski vacuum $|0  \rangle $ is precisely a thermal state in Rindler space.

The effect is rooted in the existence of the causal horizon.  The trace over the unobservable parts of the Hilbert space naturally leads to a mixed state, which for deep reasons tied to the Lorentz symmetry of the vacuum, turns out to be exactly thermal in any quantum field theory \cite{Sewell}.  The precise statement, which can be proved using path integral methods \cite{Unruh:1983ac}, is that
\beq  \langle 0|\hat{O}^{\rm R}(t,x^i,z)|0  \rangle  = Tr \left[e^{\frac{2\pi H^{\rm R}}{\kappa}}~ \hat{O}^{\rm R}(\tau,x^i,\xi)
\right]. \label{MinkowskiRindler} \eeq
Therefore, the Minkowski vacuum expectation value of any operator $\hat{O}^{\rm R}$ restricted to Rindler wedge (i.e. $z \geq |t|$) is equivalent to a thermal average at the constant Unruh-Tolman temperature $T_0 = {\kappa}/{2 \pi}$.  As said, the factor $\kappa$ is an arbitrary rescaling factor for the proper time $\tau$, as such it can always be set to one. We shall however keep it explicit for the moment as a bookkeeping quantity for the Rindler wedge temperature.

Since the vacuum is a thermal state in the Rindler wedge, one can study the entropy associated with this system.  This idea goes back to two seemingly different calculations by 't Hooft \cite{'tHooft:1984re} and Bombelli, Koul, Lee, and Sorkin (BKLS) \cite{entanglement}. 't Hooft calculated the thermal partition function at the Hawking temperature for a scalar field outside a very massive Schwarzschild black hole, which the Rindler spacetime closely approximates. Since the local Unruh temperature $T = T_0 (\kappa \xi)^{-1} = (2 \pi \xi)^{-1}$ diverges at the horizon, the entropy diverges as well and must be regularized by replacing the horizon with a ``brick wall" boundary condition. The resulting entropy scales like the cross-sectional area of the horizon.

Separately, BKLS pointed out that there is also generic statistical von Neumann entropy
\beq S_{ent} = - Tr \hat{\rho} \ln \hat{\rho}, \eeq
where $\hat{\rho}$ is an entanglement density matrix that results from tracing over the unobservable regions of the Hilbert space. This entropy exists even at zero temperature and in general scales like the area of the horizon boundary. Therefore it must be regularized with a ultraviolet (UV) cutoff, which yields in four spacetime dimensions $S_{ent} \sim A/\ell_c^2$.

For the spacetimes with a causal horizon, the thermal and quantum pictures of the entropy turn out to be equivalent because, as we saw in (\ref{MinkowskiRindler}), the density matrix $\hat{\rho}$ is precisely a thermal Gibbs state. Since a natural choice for the UV cutoff is roughly the Planck length, the entanglement entropy should be an important part of the Bekenstein-Hawking (BH) entropy. However, in general, all quantum fields will make a contribution to the entanglement entropy. This leads to the so called ``species problem":  the entanglement entropy depends on the number and type of fields, while the BH entropy is universally $A/4\ell_p^2$.

However, the entanglement entropy can be thought of as a one-loop quantum correction to the classical, tree-level BH entropy.  The quadratic divergence in the entanglement entropy seems to match the quadratic divergence that appears in the renormalization of the Newton constant \cite{Susskind:1994sm, Jacobson:1994iw, Demers:1995dq,Larsen:1995ax}. Thus, the species problem might be solved by taking the Newton constant in the BH entropy to be the renormalized Newton constant.  In this context, an attractive idea is that there is no tree-level gravitational entropy and that entanglement entropy (or in general the matter contribution to the entropy) is all the BH entropy \cite{Jacobson:1994iw, Frolov:1996aj}. This would imply the Newton constant and gravity itself is entirely ``induced" by quantum matter fluctuations, as first argued by Sakharov \cite{Sakharov:1967pk}.

\section{Rindler wedge hydrodynamics}

In this section we will review two examples where a shear viscosity emerges from the hydrodynamics of the Rindler wedge.  In the first example a global Rindler spacetime is being perturbed on a large scale with the dynamics governed by Einstein equation \cite{Eling:2009pb, Eling:2009sj}. To work conveniently at the horizon we rewrite the metric (\ref{Rindlermetric}) in Eddington-Finkelstein like coordinates with the following parametrization
\bea
v &=& \tau+(2\kappa)^{-1} \ln (r) \nonumber\\
r &=& \kappa \xi^2 \nonumber \\
\tilde{x}^i &=& \kappa^{-1} x^i,
\eea
so that the metric has the form
\beq   ds^2 = \kappa r dv^2  - dv dr  - \kappa^2 \sum^{d}_{i=1} d\tilde{x}^i d\tilde{x}_i.
\label{EFRindler} \eeq
Consider a uniform boost of the Rindler spacetime (\ref{EFRindler}) in $\tilde{x}^i$ directions, which is an isometry of the vacuum state. The result is a boosted metric
\beq ds^2 = \kappa r u_\mu u_\nu d\tilde{x}^\mu d\tilde{x}^\nu -  u_\mu d\tilde{x}^\mu dr - \kappa^2 P_{\mu \nu} d\tilde{x}^\mu d\tilde{x}^\nu, \label{boostRindler}\eeq
where the ($d+1$)-dimensional set of coordinates is $\tilde{x}^\mu = (v, \tilde{x}^i$), the ($d+1$)-dimensional vector $u^\mu = (\gamma, \gamma v^i)$ (i.e. $u^\xi = 0$), and the projection tensor $P_{\mu \nu} = \eta_{\mu \nu} + u_\mu u_\nu$. One can think of this bulk spacetime as describing a general flow of the thermal state with velocity $v^i$ with respect to the frame of a static observer.

Now imagine, for example, gravitational waves are impinging on the system. To parameterize the perturbations, we can take $u^\mu(\tilde{x}^\mu)$ and $\kappa(\tilde{x}^\mu)$ (thereby associating a scale with $\kappa$), so that the temperature and ($d+1$)-velocity of the flow are slowly varying functions of the $\tilde{x}^\mu$ coordinates. In particular, the hydrodynamic limit requires the scale $L$ of the perturbations to satisfy $L \gg \kappa^{-1}$. The metric
\beq ds^2 =\kappa(\tilde{x}) r ~ u_\mu(\tilde{x}) u_\nu(\tilde{x}) d\tilde{x}^\mu d\tilde{x}^\nu -  u_\mu(\tilde{x}) d\tilde{x}^\mu dr - \kappa^2(\tilde{x}) P_{\mu \nu} d\tilde{x}^\mu d\tilde{x}^\nu \label{boostedRindler} \eeq
is no longer flat and hence does not satisfy $R_{AB} = 0$. However we can obtain a solution (at least in principle) to the vacuum Einstein equations working order by order in a derivative expansion. We take $u(\varepsilon \tilde{x}^\mu)$ and $\kappa(\varepsilon \tilde{x}^\mu)$ where $\varepsilon$ is a book keeping factor (set to unity at the end of calculations) to keep track of derivatives of temperature and velocity. For example, at lowest order there should be solution to the equations $R_{AB} = 0 + O(\varepsilon^2)$ of the form
\beq g_{AB} = g_{AB}^{(0)} + \varepsilon g_{AB}^{(1)}(\partial u, \partial \kappa), \eeq
where $g_{AB}^{(1)}$ is a $O(\varepsilon)$ correction to the metric (\ref{boostedRindler}).

In the membrane paradigm, we want to consider the subset of $(d+1)$ vacuum Einstein equations projected into the Rindler horizon
\beq R_{\mu \nu} \ell^\nu =  0, \label{horizonproj}\eeq
where $\ell^\mu$ is the null normal to the horizon. At lowest order, $\ell^\mu=u^\mu$. Note that $u^\mu$ is unit normalized with respect to the flat metric $\eta_{\mu \nu}$, but is null on the horizon ($r=0$) of the full bulk metric.   Using the horizon Gauss-Codazzi equations and the membrane paradigm, this set of Einstein equations can be expressed solely in terms of horizon geometrical variables - i.e. the extrinsic curvature components (the horizon shear, expansion, surface gravity) and intrinsic metric of the horizon surface. At the lowest orders in $\varepsilon$, it is sufficient to calculate these quantities directly from the metric (\ref{boostedRindler}), the near-horizon data, and a choice of gauge.  For example, the horizon shear is just the fluid shear, which is given by the symmetric, trace-free transverse part of $\partial_\mu u_\nu$
\beq \tilde{\sigma}_{\mu \nu} = P^\sigma_\mu P^\tau_\nu (\partial_\sigma u_\tau + \partial_\tau u_\sigma - 2/d~ \eta_{\sigma \tau} \partial_\gamma u^\gamma) , \label{fluidshear} \eeq
and the horizon expansion is
\beq \tilde{\theta} = \partial_\mu u^\mu + d u^\mu \partial_\mu \ln \kappa. \eeq
Remarkably, up to $O(\varepsilon^2)$ the Einstein equations (\ref{horizonproj}) imply
\beq  R_{\mu \nu} \ell^\nu = \partial_\nu {T_{(F)}{}^\nu}_\mu = 0, \eeq
where $\partial_\nu {T_{(F)}{}^\nu}_\mu = 0$ are the hydrodynamic equations of a viscous conformal fluid living on a flat Minkowski metric in one less dimension. In general, a viscous fluid stress tensor has the form of a perfect fluid, plus shear and expansion terms that are first order in $\varepsilon$
\beq {T_{(F)}{}^\nu}_\mu = \epsilon u^\mu u_\nu + P(\delta^\mu_\nu + u^\mu u_\nu) - 2 \eta \sigma^\mu_\nu - \xi_B (\partial_\sigma u^\sigma) \delta^\mu_\nu. \label{fluidstress} \eeq
Here $\xi_B = 0$, consistent with the conformal condition that ${T^\mu}_\mu = 0$, while the shear viscosity is $\eta = v/16\pi G_N$ \cite{Eling:2009sj}, where $v$ is a scalar area density associated with the horizon. Assuming a Bekenstein-Hawking area entropy density $v/4G_N$, the shear viscosity to entropy density ratio turns out to be precisely the KSS ratio.

The second example is the proposal (first due to  \cite{Jacobson:1995ab} and expanded upon in \cite{RindlerKSS}) that the Einstein equations and macroscopic spacetime dynamics are just the thermodynamics of the local vacuum state. The idea is to impose a general entropy balance law on the vacuum state in Rindler wedge, since it has temperature and entropy.  One assumes a finite entropy density $s$ per unit horizon area, with $s$ possibly dependent on the field content.  When the thermal density matrix $\hat{\rho}$ at temperature $T_0$ in (\ref{MinkowskiRindler}) is perturbed, the change in entanglement entropy is related to the change in mean energy via
\beq dS = \d  \langle E  \rangle /T_0 + \delta N, \eeq
where the additional $\delta N$ is an irreversible internal entropy production term, or ``uncompensated heat".  Using linear constitutive relations between fluxes of momentum in a fluid and the thermodynamic ``forces" given by gradients of a fluid velocity, the entropy production term can be expressed in terms of the squared shear $\tilde{\sigma}_{\mu \nu}$ and expansion $\tilde{\theta}$ of the flow
\beq \delta N = \frac{2 \eta}{T_0} \tilde{\sigma}_{\mu \nu} \tilde{\sigma}^{\mu \nu} + \frac{\xi_B}{T_0} \tilde{\theta}^2, \label{eq:unheat}\eeq
where $\eta$ and $\xi_B$ are the shear and bulk viscosities respectively \cite{Landau,deGroot}.  Because the change in the mean energy is due to the flux into the unobservable region of spacetime, which is perfectly thermalized by the horizon system, it is assumed to consist entirely of heat. Thus, we have the thermodynamic entropy balance law $dS = \d Q/T_0+\delta N$.
The second assumption consists in the fact that this relation should hold for all causal horizons, with $\d Q$ as the flow of boost matter energy across the horizon.  Since the area of the horizon is no longer fixed, the spacetime must become dynamical.

In a general spacetime, a local horizon can be defined in analogy with a black hole horizon. A global
definition of the latter is the boundary of the causal past of future null infinity. The segment of a black hole horizon to the causal past of a
spatial cross-section is the boundary of the past of that cross section. A local horizon at a point $p$ is defined in a similar way: choose a spacelike 2-surface patch $B$ including $p$, and choose one side of the boundary of the causal past of $B$. Near $p$,  this boundary is a
congruence of null geodesics orthogonal to $B$, which comprises the horizon.

The equivalence principle implies the spacetime in the neighborhood of any point $p$ is approximately flat.  In this patch of spacetime one can always construct a local Rindler wedge and associate the boundary of the causal past of  $B$ with the local Rindler horizon.  At $p$ the expansion and shear of the horizon in terms of the local boost Killing vector $\chi^\mu$ automatically vanish, because $p$ is a fixed point of the local boost flow. This defines the notion of local equilibrium.

To compute the entropy change $\d S= s \d A$, one must follow the
area change of the horizon (here we assume a four dimensional bulk spacetime)
\beq \d A = \int \th \, d \lambda d^2A, \label{dA} \eeq
where $\th=d(\ln d^2A)/d\lambda$ is the expansion of the
congruence of null geodesics generating the horizon and $\lambda$ is an affine parameter along the geodesics. Using the
Raychaudhuri equation,
\beq \frac{d\th}{d\l} = -\frac{1}{2}\th^2-\s_{\m
\n}\s^{\m \n}-R_{\m \n}k^\m k^\n, \label{Ray}\eeq
the entropy change is given up to $O(\l^2)$ by the series expansion
\beq \d S = s ~ \int \left[\th
-\l \left(\frac{1}{2}\th^2+\s_{\m \n}\s^{\m \n}+
R_{\m \n}k^\m k^\n \right)\right]_p d\l  d^2A. \label{dS}\eeq
Here $k^\m = - \lambda^{-1} \chi^\mu $ is affinely parameterized tangent vector to the horizon, and $\s$ is the shear with respect to the affine flow. Note that all quantities in the integrand are evaluated at $p$.  The heat flux across the horizon has the form
\beq \frac{\d Q}{T_0} = 2\pi \int T^{M}{}_{\m \n} k^\m
k^\n (-\l) d\l d^2A, \label{dQ/T} \eeq
where $ T^{M}{}_{\m \n}$ is the matter stress tensor. If it is required that the entropy balance law holds at all points
$p$, we first find that the affine expansion at $p$ must vanish since the heat flux (\ref{dQ/T})
vanishes at $p$.  At $O(\l)$ the integrands of (\ref{dQ/T}) and
(\ref{dS}) then imply the relation
\beq (2\pi) T^{M}{}_{\m \n} k^\m k^\n =  s R_{\m \n}k^\m k^\n, \label{Clausiusfinal}\eeq
describing reversible processes, and the identification
\beq  \delta N = \frac{s}{\kappa}  \int \tilde{\sigma}_{\mu \nu} \tilde{\sigma}^{\mu \nu} d\tau d^2 A \label{HH} \eeq
in the irreversible sector, where in the above expression we have substituted the affine expansion and shear with their Killing analogues: $\tilde{\theta}=- \kappa\lambda\, \theta$ and   $\tilde{\sigma}=- \kappa\lambda\,  \sigma$.
%
%\beq
%\hat{\theta}=\left(\frac{d\lambda}{dv}\right)\theta=- \kappa\lambda\, \theta \,\,\,\,\, \mbox{and}\,\,\,\,\,  \hat{\sigma}=\left(\frac{d\lambda}{dv}\right)\sigma=- \kappa\lambda\,  \sigma, \label{si}
%\eeq
%
Now, comparison of \eqref{HH} with \eqref{eq:unheat} implies $\eta = s T_0/2\kappa = s/4\pi$, which is exactly the KSS ratio.

Furthermore, Eqn. (\ref{Clausiusfinal}) holds for all null vectors $k^\mu$. This implies
\beq R_{\m \n} + \Phi g_{\m \n} = (2\p/ s)~T^{M}{}_{\m \n}
\label{eos} \eeq
where $\Phi$ is a so far undetermined function. The free function $\Phi$ can be fixed if it is
assumed that the matter stress tensor is divergence free,
corresponding to the usual local conservation of matter energy.
Taking the divergence of both sides of (\ref{eos}) and using the
contracted Bianchi identity $\nab^\n R_{\m \n}=\frac{1}{2}\nab_\m R$ we then find that $\Phi=-\frac{1}{2}R
- \Lambda$, corresponding to the Einstein equation with
undetermined cosmological constant $\Lambda$. The derived equation describing reversible changes matches the Einstein equation, with Newton's constant determined by the entropy density $s$,
\beq G_N=\frac{1}{4s}, \eeq
or conversely, $s =1/4 G_N=1/4L_P^2$. This also implies that the shear viscosity is $1/16\pi G_N$ and that the dissipative term \eqref{HH} can be exactly identified with the well-known Hartle-Hawking formula for the tidal heating of a classical black hole~\cite{TP, HH, Chandra, Poi:2004, Poi:2005}.

It seems that once we demand a finite area entropy density for Rindler horizons, an entropy balance law can naturally imply gravity. {This would more generally indicate that} any Lorentz invariant quantum field theory with a UV cutoff (and therefore a finite entropy and a large, but finite number of degrees of freedom) must have gravity.  Interestingly, this sort of induced gravity is consistent with the AdS/CFT correspondence. In the usual formulation, the CFT on the boundary has no cutoff and infinite entanglement entropy. This corresponds to the case where $G_N^{(d+1)}=0$ and the CFT on the boundary is not coupled to gravity. Introducing a cutoff to the CFT corresponds to a brane in the AdS bulk that cuts off the region from some radial coordinate $r_0$ to infinity. The dual CFT on the brane is coupled to gravity and has a finite entanglement entropy that seems to match the BH entropy \cite{AdSent}.

\section{Microscopic description and Kubo formula}

Together the two examples above provide a mutually consistent picture of a shear viscosity coefficient emerging from large scale perturbations of the Rindler thermal state. Typically, in classical hydrodynamics the viscosities are phenomenological coefficients, either measured directly in the laboratory or calculated by matching to a microscopic description of the fluid system. However, in the above examples, our classical calculations require both the entropy density and the viscosity to have a trivial relation to the observed low energy Newton constant.  All the dependence on the number and nature of the quantum fields is apparently absorbed into this quantity.  In order to explore this unexpected universality further, we would like to find a microscopic description for the shear viscosity in terms of the fluctuations of a thermal state in a finite temperature quantum theory.

First,  it is instructive to consider calculations of viscosity in the AdS/CFT (or more generally, ``gauge/gravity") correspondence. In this case, $\eta$ and $s$ are the viscosity and entropy density of an infinitely strongly coupled ($d+1$)-dimensional finite temperature gauge theory with a dual gravitational description in terms of a black hole or brane in AdS spacetime.  The gauge theory lives in flat Minkowski spacetime and is thought of as being on the hologram at the AdS boundary.  In the duality prescription, a massless field $\phi$ in the bulk spacetime is dual to an operator $\cal{O}$ in the boundary field theory. In particular, perturbations of the bulk field act as sources for the field theory operators on the boundary via the coupling
\beq \int \phi_0 {\cal O}~ d^{d+1} x, \eeq
where $\phi_0$ is the boundary value. For small perturbations, determining the change of the expectation value of ${\cal O}$ is a well-known problem in time dependent perturbation theory. In Fourier space $(k^0, \vec{k})$ the result is \cite{Kapusta}
\beq  \langle \delta{\cal O} (k^0,\vec{k})   \rangle  = G_{R}(k^0, \vec{k}) \phi_0(k^0, \vec{k}), \eeq
where $G_R$ is the retarded two point thermal Green's function (the brackets represent a thermal average) of ${\cal O}$,
\beq G_{R}(k^0, \vec{k})  = \int dt d^d x e^{i k^0 t} e^{-i \vec{k} \cdot \vec{x}} \theta(t)   \langle [{\cal O}(x), {\cal O}(0)]  \rangle .  \eeq
On the other hand,  linear response theory \cite{deGroot} implies that in the large scale limit $k^0, \vec{k} \rightarrow 0$
\beq   \langle \delta {\cal O} (k^0,\vec{k})  \rangle  = \chi \partial_t \phi_0, \eeq
where $\chi$ is some generic phenomenological transport coefficient. Matching these two descriptions, one finds the Kubo formula
\beq \chi = \lim_{k^0 \rightarrow 0} \frac{1}{k^0} Im G_R(k^0, \vec{k}=0).  \label{simpleKubo} \eeq
Therefore, generic dissipative transport phenomena are described by fluctuations about the thermal equilibrium state. In the case of shear viscosity, the relevant field operator ${\cal O}$ is the stress tensor $T^{xy}$ (or, in general, the trace-free spatial parts of $T^{\mu \nu}$, see (\ref{fluidstress})), while the classical source $\phi$ is identified with corresponding transverse metric perturbations, for example $h_{xy}$.

The prescription for computing the retarded Green's function
is to first solve the perturbation equations for $h_{\mu \nu}$, subject to the Dirichlet condition at the asymptotic boundary and requiring at the horizon the field be purely ingoing \cite{Son:2002sd}. From the on-shell action, one can derive \cite{Iqbal:2008by}
\beq \chi =  \lim_{k^\mu \rightarrow 0} \lim_{r \rightarrow \infty} \frac{\Pi(r, k^0, \vec{k})}{i k^0 \phi(r, k^0, \vec{k})}, \eeq
where $\Pi$ is the radial canonical momentum conjugate to the field. In the low frequency limit it turns out that radial evolution of $\Pi$ is trivial. Essentially all the relevant physics is at the horizon and this is the natural place to evaluate the above quantity.
In the near-horizon limit the geometry {of a black hole solution dual to a gauge theory thermal state} reduces to the Rindler metric. Furthermore, in the membrane paradigm, the condition that fields be regular at the horizon immediately fixes the shear viscosity in terms of the coupling constant for transverse gravitons. In Einstein gravity, the result is simply the universal gravitational coupling $\eta = (16\pi G_N)^{-1}$ (which matches the results discussed in Section II), while in higher derivative theories one can derive a formula for $\eta$ in terms of horizon quantities similar to Wald's Noether charge formula for the entropy \cite{effcouplings}.

In flat Rindler space, there is no holographic duality of the AdS/CFT type, i.e. no string theoretic mapping between classical bulk fields and operators in a strongly coupled theory and no timelike boundary surface at infinity capable of supporting a dual holographic theory. Therefore the type of constructions reviewed above for calculating Green's functions do not appear to be available to us. However, as we have seen, there is a type of holography at the horizon due to entanglement when observables in the vacuum state are restricted to a subregion. For example, the entropy of fields in the Rindler wedge is naturally associated with the horizon boundary. Since the degrees of freedom in the wedge are packed into this membrane surface, the physics of the bulk spacetime can be effectively reduced to a lower dimensional description associated with a ``stretched horizon" boundary. Therefore the shear viscosity associated with the Rindler horizon must be induced by the matter fields in the quantum vacuum state, just like the entanglement entropy.

The dual lower dimensional description of the vacuum state and the near-horizon degrees of freedom are characterized by the stress-energy tensor \eqref{fluidstress} and as such can be associated to a strongly coupled thermal CFT living effectively on the flat Minkowski metric $ds^2 = \eta_{\mu \nu} dx^\mu dx^\nu = d\tau^2- \sum_i dx_i dx^i$. In addition, we expect the total energy-momentum in the bulk Rindler space should be the total energy-momentum of the dual description.

In Rindler space the explicit translational symmetry in the $z$ (or $\xi$) direction is broken. However, the symmetry in the other directions remains, so that the Lagrangian of a field theory must be invariant under
\beq x^\mu \rightarrow x^\mu + a^\mu. \eeq
Using the Noether theorem we can write a canonical energy-momentum tensor for the bulk fields in the Rindler spacetime
\beq T_{(R)}{}^\mu_\nu = \frac{\partial L_R }{\partial (\partial_\mu \psi)} \partial_\nu \psi - \delta^\mu_\nu L_R, \label{Rindlerstress} \eeq
where $\psi$ represents a generic matter field.  This stress tensor is conserved quantity in the flat spacetime sense: $\partial_\mu T_{R}{}^\mu_\nu  = 0$.  Note that the Lagrangian density is $L_R = \sqrt{-g} L_{\rm Mink}$ (where $L_{\rm Mink}$ is the field Lagrangian in Minkowski spacetime) and evaluates to $L_R = \kappa \xi L_{\rm Mink}$. Therefore, the canonical energy-momentum tensor for the Rindler wedge is $\kappa \xi$ times the $(\mu \nu)$ components of usual Minkowski space stress tensor, $T^\mu_\nu$.

On large scales, the holographic state must be described by a conserved lower dimensional stress tensor operator $ \langle \hat{T}^{(d+1)}{}^{\mu \nu}  \rangle $,
\beq \partial_\mu  \langle \hat{T}^{(d+1)}{}^{\mu \nu}  \rangle  = 0. \eeq
Here the brackets represent a thermal average $Z^{-1} Tr(\rho \hat{T}^{(d+1)}{}^\mu_\nu)$ at the Tolman-Unruh temperature, which by (\ref{MinkowskiRindler}) is equivalent to the Minkowski vacuum expectation value $ \langle 0 | \hat{T}^{(d+1)}{}^\mu_\nu| 0  \rangle $. As a simple ansatz we assume
\beq   \langle \hat{T}^{(d+1)}{}^\mu_\nu  \rangle  = \int^{\infty}_{\ell_c} d\xi \, \langle \hat{T}_{(R)}{}^\mu_\nu  \rangle  =  \int^{\infty}_{\ell_c} d\xi \, \kappa \xi  \langle \hat{T}^\mu_\nu  \rangle , \label{onepointansatz}\eeq
that the energy-momentum density in the lower dimensional description is a radial integral of the bulk quantities, which as usual must be cut off at a stretched horizon located at proper distance $\ell_c$ from the true horizon in order to be rendered finite.

This prescription is consistent with the literature on thermodynamic quantities in Rindler wedge.  The Minkowski vacuum expectation value $ \langle 0 | \hat{T}^A_B| 0  \rangle $ for free spin-0, spin-1/2 and spin-1 fields in the Rindler wedge was calculated long ago \cite{Sciama}.  To regularize the stress tensor operator, one can impose a Fulling-Rindler subtraction
\beq  \langle F|\hat{T}^A_B|F  \rangle  = 0. \label{Fullingsubtr}\eeq
As expected, one finds that the Minkowski vacuum expectation value has the form of a perfect fluid stress tensor. For example, in four spacetime dimensions the bulk energy density for a scalar field has the Planckian form
\beq \epsilon(\xi) = \frac{\pi^2 T^4}{30} = \frac{1}{480 \pi^2 \xi^4}. \eeq
From our ansatz (\ref{onepointansatz}), we find an energy density that appropriately scales like the area of the horizon boundary \cite{Dowker}
\beq \epsilon^{2+1} = \frac{\kappa}{960 \pi^2 \ell_c^2}. \eeq
Using the Gibbs relation $\epsilon+P = s T$, and equation of state $\epsilon = 3P$ for the massless bulk scalar field, we find the entropy density $s$ obeys
\beq s = \frac{2 \pi^3}{45} T^3  =  \frac{1}{180 \pi \xi^3}.  \label{volumeentropy}\eeq
Integrating over $\xi$ from $\ell_c$ to $\infty$ to find the effective area entropy yields
\beq s = \frac{1}{360 \pi \ell_c^{2}}, \label{entropycalc} \eeq
which agrees with standard results in the literature for the brick wall/entanglement entropy~\cite{'tHooft:1984re,Susskind:1994sm,Dowker}.

If we apply the formalism of viscous hydrodynamics to this system, the shear viscosity should be given by the Kubo formula (\ref{simpleKubo}) in terms of the effective stress tensor of the lower dimensional theory associated with the horizon
\beq \eta = \lim_{\omega \rightarrow 0} \frac{1}{\omega} \int d \tau d^d x e^{i \omega \tau} \theta(\tau)  \langle [T^{d+1}_{xy}(\tau,x,y), T^{d+1}_{xy}(0)]\rangle, \eeq
where $\omega$ is a Rindler frequency. Using our ansatz that the lower dimensional densities are radial integrals of the bulk matter stress-tensor, we arrive at the following formula
\beq \eta = \lim_{\omega \rightarrow 0} \frac{1}{\omega} \int^{\infty}_{\ell_c} d\xi' \int^{\infty}_{\ell_c} d\xi \int d \tau d^d x e^{i \omega \tau} \theta(\tau) \kappa^2 \xi \xi'  \langle [T_{xy}(\tau,x,y,\xi), T_{xy}(0,\xi')]\rangle. \label{RindlerKubo}\eeq
Since we have translational invariance in $(\tau,x,y)$, we can safely choose one of the points to be at $\tau=x=y=0$, so that the most general expression is a function $G^R_{xy,xy}(\tau,x,y,\xi,\xi')$. This type of expression is similar to those developed in \cite{BrusteinYarom}. The authors showed that correlation functions of certain operators expressed as an integral of a density over a sub-volume of Minkowski are UV divergent and scale like the horizon/boundary area. As an example, they found the heat capacity due to entanglement in the Rindler wedge.

As a first test case of our viscosity formula, we consider the thermal state to consist of a free, minimally coupled scalar field in a four dimensional Rindler spacetime.  One apparent problem with this choice is that the shear viscosity in an free field theory is typically ill-defined. In physical terms, shear viscosity measures the rate of transverse momentum diffusion between the elements of a fluid. Although the quasi-particle description in kinetic theory is not a good one in a strongly coupled system, we can gain some guidance by thinking of shear viscosity as a diffusion process. One can show that $\eta \sim \epsilon l_{\rm mfp}$, where $l_{\rm mfp}$ is the mean free path of the fluid. Since in a free field theory the mean free path diverges, $\eta$ diverges as well. This is just a consequence of the breakdown of the effective hydrodynamic theory.

On the other hand, in our case the equivalence principle implies a field theory in Rindler space can be thought of as being in a constant gravitational field. As we argued in Section II, imposing a UV cutoff on this system seems to introduce gravitational dynamics. If the cutoff is placed near the Planck length (as we suspect) the gravitational dynamics is strongly coupled there. The idea is that the dominant effect in the relaxation of the vacuum thermal state is the strongly coupled gravitational interaction. This also seems to explain how there can be universality in the result for $\eta$. In principle, all quantum matter fields should be present in the the vacuum state. However, the ratio $\eta/s$ should be $1/4\pi$ regardless of the type of quantum fields in the wedge or the dimension of the spacetime. Since gravity interacts with all fields in the same way, it should not make a difference whether we consider the soup of fields to be made up of a free scalar field, free fermions, or some type of interacting fields.

\section{Viscosity calculation}

Since the thermal average is at the Tolman-Unruh temperature $T_0$, by (\ref{MinkowskiRindler}) it is equivalent to an ordinary Minkowski vacuum expectation value
\beq  \langle 0 |[T_{xy}(\tau,x,y,\xi), T_{xy}(0,\xi')]|0\rangle \eeq
which makes calculations much simpler.  One can compute the correlator in the Minkowski vacuum state, change from inertial coordinates $X^A$ to Rindler coordinates $Y^A$ and then perform the Fourier transform.  The Minkowski stress tensor for a free, massless scalar field has the form
\beq T^A_B = \frac{\partial L }{\partial (\partial_A \phi)} \partial_B \phi - \delta^A_B L, \label{canonicalstress} \eeq
where $L = g^{AB} \partial_A \phi \partial_B \phi$. One can insert this in Eqn. (\ref{RindlerKubo}) which is in terms of the retarded Green's function, but it is also possible to write the Kubo formula in terms of different types of Green's functions.  Since the thermal Green's functions satisfy the relation \cite{Son:2002sd}
\beq G^{1}(\omega, \bm{p}) = -\coth\left(\frac{\omega}{2T}\right) Im G^{R}(\omega, \bm{p}), \label{Greenrelation0} \eeq
where the $G$'s represent Green's functions constructed from any local bosonic operator, one can also work with the symmetrized Schrodinger-Hadamard correlator of the stress tensor $G^{1}(\omega, \bm{p})$ . So we have, for example
\beq \eta = \frac{1}{2T_0} \lim_{\omega  \rightarrow 0}  \int^{\infty}_{\ell_c} d\xi' \int^{\infty}_{\ell_c} d\xi  \int e^{i \omega \tau} dt \int d^2 x\, \kappa^2 \xi \xi' G^{1}_{xy,xy}(\tau,x,y,\xi,\xi'). \label{RindlerKubo2} \eeq
Furthermore, in the hydrodynamic limit ($\omega, \bm{k} \ll \hbar^{-1} T_0$) the symmetrized correlator is not different from the Wightman correlator
\beq G^{+}_{xy,xy} =  \langle\hat{T}_{xy}(\tau,x,y,\xi,\xi') \hat{T}_{xy}(0,0,0,\xi')\rangle. \eeq
At the quantum level the difference between the correlators in frequency space is smaller than the correlators themselves by the factor $ \omega/T$, and the hydrodynamic limit here is exactly where $\omega \ll T$ \cite{Kovtun:2003vj}.

In practice, we found it was easiest to work with the Wightman correlator. We first expand the scalar field operator into the usual set of normal mode solutions to the Klein-Gordon field equation
\beq \hat{\phi}(t,\bm{x}) = \int \frac{d^{d+1} p}{(2\pi)^{d+1} \sqrt{2\omega}} \left[a(\bm{p}) e^{i \bm{p} \cdot \bm{x} - i \omega t} + a^{\dag}(\bm{p}) e^{-i \bm{p} \cdot \bm{x} + i \omega t}\right], \label{normalmodes}\eeq
where $\omega = |\bm{p}|$ and $a(\bm{p})$ and $a^{\dag}(\bm{p})$ are creation and annihilation operators.  Inserting this into the Wightman function, we find
\beq G^{+}{}_{xy,xy}(t,x,y,z,z') = \int \frac{d^3 p d^3 q d^3 p' d^3 q'}{4 (2\pi)^{12} \sqrt{p p' q q'}} p_x q_y p'_x q'_y  \langle0| \cdots | 0\rangle \eeq
where the $\cdots$ represent sixteen terms involving combinations of four creation and annihilation operators and exponentials of the momenta. However, the only two terms that contribute are
\bea  \langle 0 |a(\bm{p}) a^{\dag}(\bm{q}) a(\bm{p'}) a^{\dag}(\bm{q'})|0\rangle e^{-i (P_\mu-Q_\mu) x^\mu} e^{-i (P'_\mu-Q'_\mu) x'^\mu}\nonumber \\ +
 \langle 0 |a(\bm{p})a(\bm{q})a^{\dag}(\bm{p'})a^{\dag}(\bm{q'})|0\rangle e^{-i (P_\mu+Q_\mu) x^\mu} e^{i (P'_\mu+Q'_\mu) x'^\mu}, \label{creationexp} \eea
where $P_\mu = (|\bm{p}|, \bm{p})$ and $x'^\mu = (0,0,0,z')$. Using the commutation relation
\beq [a(\bm{p}), a^{\dag}(\bm{p'})] = (2\pi)^{d+1} \delta^{d+1}(\bm{p}-\bm{p'}), \label{commutation}\eeq
we find that
\beq  \langle 0 |a(\bm{p})a^{\dag}(\bm{q})a(\bm{p'})a^{\dag}(\bm{q'})|0\rangle = (2\pi)^6 \delta^{3}(\bm{p'}-\bm{q'}) \delta^{3}(\bm{p}-\bm{q}) \eeq
and
\beq  \langle 0 |a(\bm{p})a(\bm{q})a^{\dag}(\bm{p'})a^{\dag}(\bm{q'})|0\rangle =  (2\pi)^6 \left( \delta^{3}(\bm{p}-\bm{p'}) \delta^{3}(\bm{q}-\bm{q'})+ \delta^{3}(\bm{q}-\bm{p'}) \delta^{3}(\bm{p}-\bm{q'}) \right). \eeq
Putting this together and integrating gives
\beq G^{+}{}_{xy,xy}(t,x,y,z,z') = \int \frac{d^3 p d^3 q}{4 (2\pi)^{6}} \frac{1}{pq} \left[  (p_x^2 q_y^2+p_x p_y q_x q_y) e^{- i (P^\mu + Q^\mu)(x_\mu - x'_\mu)} + p_x p_y q_x q_y \right]. \eeq
The last term, which comes from the first piece in (\ref{creationexp}) seems to give an infinite contribution in general. It is associated with the summation over the zero point modes and would be absent if we had followed the usual prescription of normal ordering the stress tensor operator so that its expectation value is set to zero. Instead, if we use the Fulling-Rindler subtraction (\ref{Fullingsubtr}) these Casimir type terms must be present. In the present case, however, the integration over this term over the momentum space gives zero identically. {This is consistent with the perfect fluid form of expectation value of $\hat{T}_{\mu \nu}$, whose $(xy)$ components are zero in the equilibrium rest frame.}

In order to deal with the remaining term, note that the Wightman function for the scalar field operator is
\beq G^{+}(t,x,y,z,z') =  \langle 0|\phi(t,x,y,z)\phi(0,0,0,z')|0\rangle = \int \frac{d^3 p}{2p (2\pi)^{3}} e^{-i P_\mu (x^\mu-x'^\mu)}. \eeq
Therefore the Wightman function of the stress tensor can be expressed in terms of derivatives of the scalar field Wightman function
\bea G^{+}{}_{xy,xy}(t,x,y,z,z')  = (\partial^2_x G^{+}(t,x,y,z,z'))  (\partial^2_y G^{+}(t,x,y,z,z'))  \nonumber \\ + (\partial_x \partial_y G^{+}(t,x,y,z,z')) (\partial_x \partial_y G^{+}(t,x,y,z,z')). \label{stressfromscalar}\eea
The scalar Wightman function for a massless field has the form  \cite{BogShirkov}
\beq \frac{-1}{4\pi^2}\frac{1}{\lambda+ \epsilon(t) i \epsilon} \label{Wightman}\eeq
where $\lambda = -t^2+x^2+y^2+(z-z')^2$ is the spacetime interval between the two points and $\epsilon(t)$ is sign function (+1 if $t>0$, -1 if $t < 0$). The  $i \epsilon$ prescription for dealing with the singularity here is interpreted as
\beq \lim_{\epsilon \rightarrow 0} \frac{1}{\lambda \pm i \epsilon} = P/\lambda \mp i \pi \delta(\lambda), \eeq
where $P$ represents the Cauchy principal value.  Using (\ref{stressfromscalar}) we find the Wightman function for the stress tensor is
\beq G^{+}_{xy,xy}(\tau,x,y,\xi,\xi') = \frac{1}{16\pi^4}~\left(\frac{128 x^2 y^2}{(\lambda+i\epsilon)^6}-\frac{16 x^2}{(\lambda+i\epsilon)^5}-\frac{16 y^2}{(\lambda+i\epsilon)^5}+\frac{4}{(\lambda+i\epsilon)^4}\right) \label{coordinateHadamard}\eeq
where we have used (\ref{transform}) to re-express the interval in Rindler coordinates: $\lambda = \xi^2-2\xi \xi' \cosh(\kappa \tau)+\xi'^2+x^2+y^2$.

We want to calculate the Fourier transform into Rindler frequency and momentum (taking the zero momentum limit)
\beq \tilde{G}^{+}_{xy,xy}(\omega,\xi,\xi') = \int^{\infty}_{\ell_c} d \xi'  \int^{\infty}_{\ell_c} d \xi \int^{\infty}_{-\infty} e^{i \omega \tau} d\tau \int^{\infty}_{-\infty} dx \int^{\infty}_{-\infty} dy ~\kappa^2 \xi \xi' ~G^{+}_{xy,xy}(\tau,x,y,\xi,\xi'). \label{FourierWightman}\eeq
We first make a coordinate change to
\bea x &=& \rho \cos(\theta) \\
y &=& \rho \sin(\theta) \label{converttospherical}
\eea
so that the integrations over the $x$ and $y$ directions become
\beq \int^{\infty}_{0} \rho d\rho \int^{2\pi}_{0} d\theta. \eeq
After integrating over the angular direction (\ref{FourierWightman}) becomes
\bea \tilde{G}^{+}_{xy,xy}(\omega,\xi,\xi') &=& \int^{\infty}_{\ell_c} d \xi'  \int^{\infty}_{\ell_c} d \xi \int^{\infty}_{-\infty} e^{i \omega \tau} d\tau \int^{\infty}_{0} d\rho \kappa^2 \xi \xi' \left( \frac{2 \rho^5}{\pi^3 (\rho^2+\alpha)^6} - \frac{2 \rho^3}{\pi^3 (\rho^2+\alpha)^5} + \right. \nonumber\\ && \left. \frac{\rho}{2 \pi^3 (\rho^2+\alpha)^4} \right), \label{int1} \eea
where $\alpha = \xi^2+\xi'^2-2\xi \xi' \cosh(\kappa \tau)+i\epsilon$. Since $\alpha$ is complex valued, the integration of $\rho$ over the real axis is well-defined and yields
\beq \tilde{G}^{+}_{xy,xy}(\omega,\xi,\xi') = \int^{\infty}_{\ell_c} d \xi'  \int^{\infty}_{\ell_c} d \xi \int^{\infty}_{-\infty} e^{i \omega \tau} d\tau \frac{1}{30 \pi^2}~ \frac{\kappa^2 \xi \xi'}{(\xi^2+\xi'^2-2 \xi \xi' \cosh(\kappa \tau))^3}. \label{tauint} \eeq
This function has a periodicity in the $\tau$ coordinate due to the $\cosh$ function. We need a prescription for dealing with the poles, which are always on the real axis at
\beq \tau_0 = \pm \kappa^{-1} \ln(\xi/\xi'). \label{realpoles}\eeq
The usual way for handling these types of integrations is to assume $\tau$ is a complex variable and that the contour for the integration in the complex $\tau$ plane should be rectangular. One horizontal piece is along the real axis (the part we want), the other in the opposite direction at $\tau = i 2\pi/\kappa$ to keep the $\cosh$ function invariant. Because of this fact there are also poles at
\beq \tau_0 = \pm \kappa^{-1} \ln(\xi/\xi') + 2\pi i/\kappa. \label{imaginarypoles} \eeq
The vertical parts are at $\tau = i \infty$ and do not contribute. The result will be the sum of the residues enclosed in the contour,
\beq  I = 2\pi i (1-e^{-2\pi \omega/\kappa})^{-1} ~  \Sigma (res) , \label{contourint}\eeq
where $I$ is the integral in (\ref{tauint}).

There are multiple choices we can make for this contour depending on which poles we choose to enclose. However, it turns out we have to include an even number of the poles (two or all four) in order to preserve the symmetry of the integrand under the interchange of $\xi$ and $\xi'$.  In the case of the Wightman function, we have the explicit $i \epsilon$ prescription, which is to include both poles on the real axis in the contour (by pushing them up), while leaving out the ones at $2\pi i/\kappa$. Computing the residues, and taking the $\omega \rightarrow 0$ limit, we find
\beq \tilde{G}^{+}_{xy,xy}(0,\xi,\xi') = \frac{\xi \xi' \kappa}{30 \pi^3}~ \frac{-3(\xi^4-\xi'^4)+ 2\xi^4 \ln(\xi/\xi') + 8\xi^2 \xi'^2 \ln(\xi/\xi') + 2\xi'^4 \ln(\xi/\xi') }{(\xi^2-\xi'^2)^5}. \label{Gtauxy}\eeq

Next, we must perform the radial integrations over $\xi$ and $\xi'$. The first integration of (\ref{Gtauxy}) over $\xi$ gives
\beq \tilde{G}^{+}_{xy,xy}(0,\xi') = \frac{\kappa}{240\pi^2}  \frac{\xi'^4+4 \ell_c^2 \xi'^2-5 \ell_c^4+4 \ell_c^4 \ln(\ell_c/\xi')+8 \ell_c^2 \xi'^2 \ln(\ell_c/\xi')}{(\ell_c^2-\xi'^2)^4}. \eeq
Integrating this expression over $\xi'$ and multiplying by the overall $(2T_0)^{-1} = \pi/\kappa$ in (\ref{RindlerKubo2}), we ultimately arrive at
\beq \eta = \frac{1}{1440 \pi^2 \ell_c^2}, \label{eta}\eeq
which, as expected, is divergent in the limit $\ell_c \to 0$ and scales in $\ell_c$ as a $2+1$ quantity.

The final task is to compare this result with the entanglement entropy density for the wedge.  Comparing with our $\eta$ in (\ref{eta}) with the entanglement entropy density calculated previously (\ref{entropycalc}), we find
\beq \eta/s = 1/4\pi. \eeq
The UV cutoff length cancels out and we are left with exactly the KSS ratio.

\section{Discussion}

In this paper we have argued that the universal shear viscosity to entropy density ratio of $1/4\pi$ is also associated with a Rindler causal horizon in a flat (either globally or locally) spacetime. Its appearance in this case is mysterious since there is no gravity and the familiar formalism of AdS/CFT holography is completely absent. In order to provide a microscopic basis for this result, we have turned to the properties of quantum entanglement and vacuum fluctuations. Namely, when a quantum state is restricted to a sub-region of the spacetime (in this case Minkowski vacuum state in the Rindler wedge), quantum fluctuations of this state have a dual, thermal description associated with the horizon boundary. An effective description of the large-scale dynamics of this vacuum thermal state is always provided by hydrodynamics. To this end, we have developed a simple Kubo-like formula for the viscosity induced on the horizon in terms of a two point stress-energy tensor correlation function for the quantum fields in the Rindler wedge. We calculated this quantity in the simplest case of a free massless scalar field in a four dimensional spacetime and found the ratio of our $\eta$ to the entanglement entropy $s$ is exactly $1/4\pi$\footnote{Note that, strictly speaking, we have not proven our result is independent of the regularization scheme used on  $\eta$ and $s$. However, we do not expect the choice of regularization to matter since both divergences arise in the same radial integration over the local energy-momentum density.}.

Our result suggests that the $1/4\pi$ ratio might be a fundamental property of quantum entanglement and its associated holography. It also provides support for the hypothesis that semi-classical gravity on macroscopic scales is induced or emergent as an effective theory of some lower dimensional, strongly coupled quantum system with a large number of degrees of freedom. In this picture, the $1/4\pi$ ratio is saturated in gauge theories with an Einstein gravity dual because 1) they have an area (BH) entropy and 2) as we mentioned at the end of Section III, in the large $N$ limit the number of degrees of freedom diverges and gravity is turned off as the Newton constant goes zero.

It would be useful to understand if our results can be extended to more general quantum field theories and to higher dimensional spacetimes. Since all fields in nature contribute in principle to the vacuum fluctuations, our hypothesis is that the $\eta/s$ ratio is $1/4\pi$ universally for any matter field. Also,  the arguments of Section III can be extended to any dimension; since the BH entropy density is $(4G^D_N)^{-1}$ and $\eta = (16\pi G^D_N)^{-1}$ for a general spacetime dimension $D$, the ratio should not depend on the number of dimensions.\footnote{In the old membrane paradigm of Damour, Price, and Thorne, the ratio of the shear viscosity coefficient to entropy density is indeed $1/4\pi$ regardless of dimension. This is because the Hartle-Hawking tidal coefficient is $1/16\pi G^D$ (otherwise independent of dimension), while s is always $1/4G^D$. Note however that one has to be careful when identifying this ratio with the KSS ratio, since in the Schwarzschild spacetime a hydrodynamic long wavelength expansion is not possible.}

As a simple first check of a different field theory, we considered a massless, but now non-minimally coupled scalar field given by the action
\beq I_s = \frac{1}{2} \int \sqrt{-g} (\nabla_A \phi \nabla^A \phi - \xi_C R \phi^2). \eeq
In the flat spacetime limit, the stress tensor reduces to
\beq T_{AB} =  \partial_A \phi \partial_B \phi - \frac{1}{2} \eta_{AB} (\partial \phi)^2 - 2 \xi_C \partial_A( \phi \partial_B \phi) + 2 \xi_C \eta_{AB} \nabla_C (\phi \nabla^C \phi). \eeq
Repeating the steps at the beginning of Section IV, we arrive at the following for the Wightman function of the stress tensor $G^{+}_{xy,xy}(\tau,x,y,\xi,\xi')$ in terms of the scalar field Wightman function $G(\tau,x,y,\xi,\xi)$
\bea G^{+}_{xy,xy}(\tau,x,y,\xi,\xi') = (1-2\xi_C)^2 (\partial^2_x G^{+}) (\partial^2_y G^{+}) - 4\xi (1-2\xi) (\partial_x G^{+}) (\partial_x \partial^2_y G^{+}) \nonumber \\ - 4\xi (1-2\xi) (\partial_y G^{+}) (\partial^2_x \partial_y G^{+}) + 4\xi_C^2 G^{+} (\partial^2_x \partial^2_y G^{+}) + (1-4\xi_C+8\xi_C^2) (\partial_x \partial_y G^{+}) (\partial_x \partial_y G^{+}). \eea
Inserting in the form of the Wightman function (\ref{Wightman}), we can calculate the Fourier transform in (\ref{FourierWightman}). Integrating over $x$ and $y$ as before, we find (\ref{int1}) again. The dependence on the coupling to the scalar curvature $\xi_C$ vanishes in the low momentum regime, and therefore $\eta$ is not changed.

There are different results in the literature for the entropy density $s$ of a non-minimally coupled scalar field. In \cite{Demers:1995dq}, the authors worked in the brick-wall approach, calculating the density of states for a thermal field outside the horizon. In this case, since the scalar curvature on the background spacetime is always zero, the $\xi_C$ dependence drops out of the scalar field equation and the entropy density is unchanged. This is consistent with our calculation and, if we use this result, the ratio is preserved. However, there is an important difficulty here that cannot be overlooked. The divergence in the entropy density found by \cite{Demers:1995dq} cannot be absorbed into the renormalization of the Newton constant, which is $\xi_C$ dependent. This is a problem since we have argued both the entanglement entropy and the viscosity are proportional to the (renormalized) Newton constant.

On the other hand, the entropy can also be calculated in an Euclidean functional integral approach from the one-loop effective action. In this case, one works off-shell and includes the contributions of manifolds where $\beta \neq 2\pi/\kappa$. When the solution is not the Hartle-Hawking instanton ($\beta= 2\pi/\kappa$), the manifolds have a conical singularity and therefore the scalar curvature coupling contributes a delta function term to the partition function. The resulting entropy is $\xi_C$ dependent and can in fact be reabsorbed into the renormalized Newton constant \cite{Solodukhin:1995ak, Larsen:1995ax}.  However, there is an interpretational issue with this result. When $\xi_c > 1/6$ the statistical mechanical contribution to the entropy seems to be negative, while $S_{ent} = - Tr \hat{\rho} \ln \hat{\rho}$ must be positive definite \cite{Hotta:1996cq}.

The reason for this unusual behavior is rooted in the fact that the black hole entropy has in this case an additional non-statistical term proportional to the integral of $\phi^2$ over the horizon
\beq S_{N} = 2\pi \xi_c \int_{H} \phi^2 \sqrt{-h} d^2 x, \label{Noethercorrection}\eeq
which can be thought of as a Noether charge correction term \cite{Frolov:1996aj}.
Indeed, if one considers a non-minimal field coupled to gravity
\beq I_{grav} =  \int d^4 x \sqrt{-g} \left(\frac{R}{16\pi G} + \frac{1}{2} (\nabla \phi)^2 - \frac{\xi_C}{2} R \phi^2 \right) \eeq
the resulting theory is a scalar-tensor theory of gravity, whose classical Wald Noether charge entropy \cite{Wald:1993nt} includes the correction (\ref{Noethercorrection}). Note that this kind of correction is not limited to non-minimally coupled scalar fields. It also appears in generic vector field theories \cite{Frolov:1998ea}.  The black hole entropy is generally composed of three contributions: a statistical entanglement entropy, the non-statistical ``bare" gravitational entropy and the Noether charge term. Induced gravity models remove the need for the bare gravitational entropy, but they currently cannot fully explain the existence of the Noether charge term from a statistical point of view \cite{Fursaev:2004qz}.

Hence our preliminary investigation of the viscosity to entropy density ratio in different field theories has lead us to a key issue. Namely, while the $\eta/s$ ratio seems to remain $1/4\pi$ if we compare our entanglement viscosity only to the statistical entanglement entropy, in general the relevant quantities are the black hole (Wald) entropy and likely a corresponding general definition of viscosity. The problem is that the Wald entropy in general diffeomorphism invariant theory of gravity does not just depend on the horizon area. This does not seem to fit with the induced gravity scenario implied by the thermodynamics of spacetime argument, reviewed in Section III, where the horizon entropy is purely due to entanglement.

{In this sense, the investigations \cite{RindlerKSS} seem to lend some insight towards a possible resolution. In particular}, different formulations of the equivalence principle and their role in determining the characteristics of a gravitational theory may be the {key} point. The only known theory of gravity consistent with the {\it strong} equivalence principle is Einstein gravity. The strong equivalence principle implies that gravity is purely geometrical. Physics (gravity included) is the same in any locally flat region of spacetime, which means $G_N$ is a universal constant and there are no extra gravitational fields. Under these conditions the UV cutoff $\ell_c$ should be a constant. However, in a general theory of gravity (such as scalar-tensor theories), the strong equivalence principle is not satisfied. {Consequently, it is reasonable to assume the UV cutoff  to be dependent on the spacetime location. In this case it is necessary to promote it to a spacetime field, which will have to be a dynamical one in order to assure the background independence of the resulting gravitational theory. This is exactly what is naturally suggested by the extensions of the spacetime thermodynamics approach beyond General Relativity \cite{RindlerKSS}.} If this is {true}, it may be always possible to re-express the Wald entropy in the form of an entanglement entropy by suitably characterizing the spacetime dependence of $\ell_c$\footnote{A related proposal can be found in \cite{waldentang}, where the authors found that Wald entropy evaluated on static, spherically symmetric black hole solutions in generalized theories of gravity can be expressed as $A/4G_{\rm eff}$, where $G_{\rm eff}$ is an effective gravitational coupling at the horizon.}. We leave this for future investigation.

We now move on to the viscosity to entropy density ratio in higher dimensions. To test the ratio here we considered a free scalar field in higher even dimensional Rindler spacetimes. This is again the simplest case, because in any odd spacetime dimension the scalar and stress tensor Wightman functions are inverse fractional powers of spacetime interval. In these cases the Fourier transform (\ref{FourierWightman}) has branch cuts, not just simple poles and the entire analysis has to be re-done. Here we present the calculation in six dimensions.The scalar Wightman function has form
\beq G^{+}(\tau,x,y,\xi,\xi') = \frac{1}{4\pi^3} \frac{1}{(\lambda+\epsilon(t) i \epsilon)^2}. \eeq
This form can be inserted into (\ref{stressfromscalar}); otherwise the Kubo formula is unchanged. In the Fourier transform we must now integrate over two additional transverse spatial directions using the generalization of (\ref{converttospherical}). The result is
\beq   \tilde{G}^{+}_{xy,xy}(\omega,\xi,\xi') = \int^{\infty}_{\ell_c} d \xi'  \int^{\infty}_{\ell_c} d \xi \int^{\infty}_{-\infty} e^{i \omega \tau} d\tau \frac{3}{140 \pi^3}~ \frac{\kappa^2 \xi \xi'}{(\xi^2+\xi'^2-2 \xi \xi' \cosh(\kappa \tau))^4}   .\eeq
Repeating the contour integration over $\tau$ and the integrations over $\xi$ and $\xi'$, we find that the viscosity is
\beq \eta = \frac{1}{33600 \pi^3 \ell_c^4}. \eeq
To compute the entropy density, we need the Planckian energy density for a gas of bosons in six-dimensions. This has the form
\beq \epsilon = \frac{2 \pi^{5/2}}{(2\pi)^5 \Gamma(5/2)}~ \Gamma(6) \zeta(6) T^6,\eeq
where $\Gamma(x)$ is the gamma function and $\zeta(x)$ the Riemann zeta function. Our ansatz (\ref{onepointansatz}) gives the lower dimensional energy density
\beq \epsilon^{4+1} = \frac{\kappa}{24192 \pi^3 \ell_c^4}.\eeq
Using the Gibbs relation and the equation of state $\epsilon=5P$, we find an entropy density
\beq s= \frac{1}{10080\pi^2 \ell_c^4} \label{higherdentropy} ,\eeq
which means an $\eta/s$ ratio of $3/10\pi$.

It is fairly straightforward to extend the above calculation to even higher dimensions (eight and ten) by modifying the Wightman function and the Planckian energy density. These results show that in higher dimensions the ratio is a rational number factor ($>1$) times $1/4\pi$. If there is not a mistake in our calculation, one could worry that the appearance of the KSS ratio in four dimensions is simply an amazing coincidence.

However, this seems unlikely to us. One possibility is that in higher dimensions the divergence of the entropy can only be absorbed into the renormalized Newton's constant up to an overall factor. While this is not the most likely explanation, so far we have found no literature that conclusively addresses this question. Another issue is that the energy density for the scalar field may not be purely Planckian in a higher dimensional Rindler space. This is not out of the realm of possibility since, for example, it is known that higher spin fields in Rindler do not have a Planckian form even in four dimensions \cite{Sciama}. Furthermore, the vacuum expectation value in Rindler space is related to the conformal anomaly present for fields on the conformally related metric of the Einstein universe \cite{Candelas:1978gf}. In different spacetime dimensions the conformal anomaly takes on different forms. In any case, it is conceivable our result for the entropy density (\ref{higherdentropy}) could be incorrect.

Another more subtle possibility is that the naive extension of our simple ansatz for the shear viscosity is not valid in higher dimensions. One could imagine, based on the membrane paradigm \cite{Eling:2009pb,Eling:2009sj} that the entropy density and the viscosity of the Rindler horizon are in fact controlled by a ($d+1$)-dimensional CFT associated with the near-horizon degrees of freedom.  The idea of identifying the near-horizon degrees of freedom of a black hole with a CFT has appeared many places in the literature \cite{nearCFT}, but in the past the CFT has been thought of as being universally 1+1 dimensional and identified with physics in the radial ($\xi$)-time ($\tau$) plane. Perhaps our simple ansatz for Rindler space holography matches a more fundamental description only in the special case of a four dimensional spacetime, while in higher dimensions it would not correctly reproduce the properties of the near-horizon theory. However, this is all very speculative. Making these remarks more concrete will be subject for future research.

\section*{Acknowledgments}
The authors wish to thank Y. Oz, A. Roura, and M. Visser for useful comments.

\end{document}